\documentclass[aps,twocolumn,showpacs]{revtex4}
\usepackage{epsfig}
\usepackage{psfrag}
\newcommand{\be}{\begin{equation}}
\newcommand{\ee}{\end{equation}}
\newcommand{\bea}{\begin{eqnarray}}
\newcommand{\eea}{\end{eqnarray}}
\newcommand{\ba}{\begin{array}}
\newcommand{\ea}{\end{array}}

\begin{document}

\title{Extracting Phases from Aperiodic Signals}

\author{Alexander Kraskov, Thomas Kreuz, Ralph G. Andrzejak, Harald 
St\"ogbauer, Walter Nadler and Peter Grassberger}
\affiliation{Complex Systems Research Group, John-von-Neumann Institute for 
Computing,
J\"ulich Research Center, D-52425 J\"ulich, Germany}

\date{\today}

\begin{abstract}
We demonstrate by means of a simple example that the arbitrariness of defining
a phase from an aperiodic signal is not just an academic problem, but is
more serious and fundamental. 
Decomposition of the signal into components with positive phase velocities 
is proposed as an old solution to this new problem.
\end{abstract}


\maketitle

There are some quantities which, although in principle exactly and uniquely 
defined, are hard to estimate from experimental data. Prominent examples are 
information theoretic quantities like algorithmic complexities (for which only
upper bounds can be estimated \cite{li}), Shannon entropies \cite{beirlant}, 
and mutual entropies \cite{kraskov}.
Opposed to these are quantities where even a unique definition is lacking,
although most researchers believe that they have a decent common sense 
``definition". Maybe the most important quantity in this category is the 
phase of a non-periodic signal \cite{vakman,boashash,cohen,hahn}.

For a pure sine wave signal the notion of phase is obvious and trivial, 
provided
the sampling rate is high enough (which we will assume throughout the 
following).
Things are nearly as clean for anharmonic periodic signals. There, we can 
map the orbit, e.g. by delay embedding, onto a closed loop in a plane, and we 
can 
define the phase by the angle of the vector from some point in the interior of 
the 
loop to the point corresponding to the actual state. This phase will of course
depend on the central point and on the delay, but different choices will give
equivalent phases, if the loop does not intersect itself: They will give the 
same average angular velocity $\omega = 
\lim_{T\to\infty}(\phi(t+T)-\phi(t))/T$, and the difference between two phases
defined that way will stay bounded with time. This ``geometric" definition of 
phase can be generalized to aperiodic signals whenever there is a way -- 
either
via embeddings or using multivariate time series -- to project the orbit into
a plane in such a way that it always encircles some point. A typical example 
is the R\"ossler attractor \cite{pikovsky} for particular values of its 
parameters. But if the loop intersects 
itself such that its interior is divided into several 
domains, then central points chosen in different regions will lead to 
non-equivalent phases (see Fig.1)

\begin{figure}
  \begin{center}
   \psfig{file=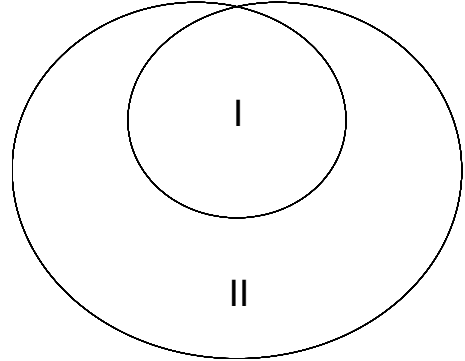,width=4.8cm, angle=270}
   \caption{Projection of a periodic orbit into a plane such that it
     intersects itself and gives rise to 2 domains. For central points 
     chosen in domain II the mean angular velocity is half of that for 
     central points in domain I.}
  \end{center} 
\end{figure}

In addition to this geometrical definition, another popular approach is 
via the Hilbert transform \cite{pikovsky,vakman,boashash}. Under mild 
restrictions on the signal $x(t)$, the pair $\{x(t),y(t)=(Hx)(t)\}$ 
form real and imaginary part of an analytic function, and the phase 
is defined as its argument in the complex plane. In general this 
gives a well-defined phase. Its value changes of course if the signal 
is shifted, $x(t) \to x(t)+c$, but a unique phase is obtained after 
de-meaning, i.e. when $c$ is such that $\langle x(t) \rangle = 0$ 
after the shift. Problems arise if the trajectory goes through the origin,
$x(t)=(Hx)(t)=0$ at some time $t$ \cite{cohen}. Practically the method 
breaks down even when this is not exactly fulfilled, since then the 
phase is very sensitive to low amplitude noise.

A third situation where one might be tempted to see a ``natural" way
of defining a phase is when a signal arises from circular motion in some 
$d$-dimensional space with constant amplitude $A$ and arbitrarily 
changing $\phi(t)$, 
\be
   x(t) = A \cos(\phi(t))
\ee
Certainly this is considered by many as the prototype of a situation
where a phase is uniquely defined in an obvious way. 

We now ask ourselves whether all three approaches give in general
the same phase. If this is not the case, then each approach might
be useful by itself, but it cannot claim to be really fundamental and 
universal.

In Fig.2 we show part of a signal. A phase portrait obtained by
plotting $x(t)$ against $x(t+\tau)$ is shown in Fig.3, a similar 
phase portrait using the Hilbert transform in Fig.4. In neither 
case one sees a point around which the orbit circles, thus neither
allows a clear and robust definition of phase. The spectrum, 
obtained with a Welch window, is shown in 
Fig.5. A prominent peak is seen, but this peak is not sharp
and thus a unique angular frequency seems not obtainable. One 
can try several other methods popular in signal analysis, but 
we argue that none of them will lead to a robust determination of 
a phase.

\begin{figure}
  \begin{center}
   \psfig{file=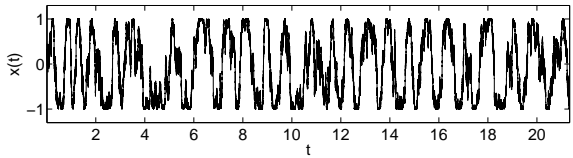,width=2.1cm, angle=270}
   \caption{Part of the signal analyzed in Figs.~3 to 5.}
  \end{center}
\end{figure}

\begin{figure}
  \begin{center}
   \psfig{file=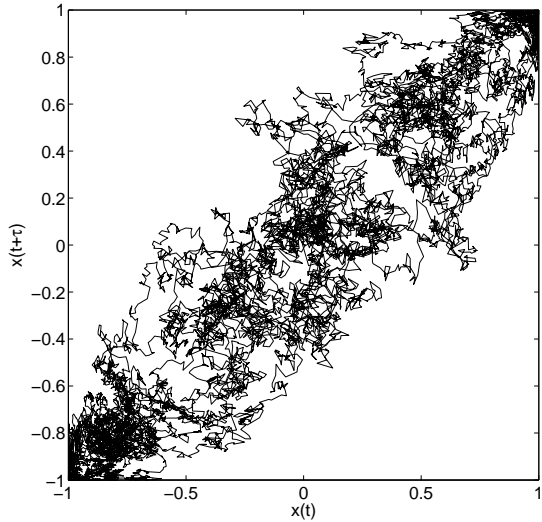,width=7cm, angle=270}
   \caption{Phase portrait of the signal shown in Fig.2, 
      obtained by plotting $x(t)$ against $x(t+\tau)$ with $\tau=0.015)$.}
  \end{center}
\end{figure}

\begin{figure}
  \begin{center}
  \psfig{file=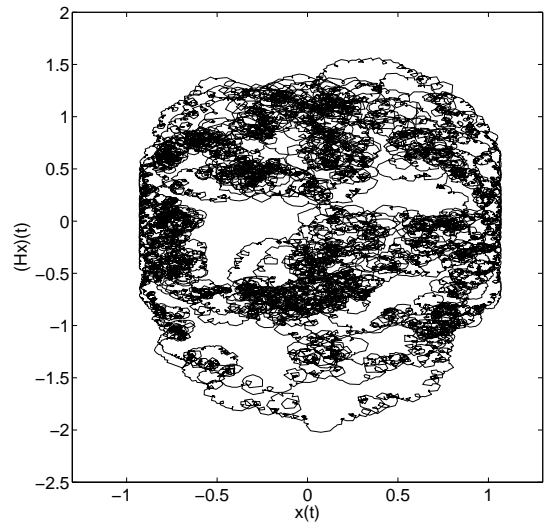,width=7cm, angle=270}
   \caption{Phase portrait of the signal shown in Fig.2,
     obtained by plotting $x(t)$ against its Hilbert transform.}
  \end{center}
\end{figure}

\begin{figure}
  \begin{center}
   \psfig{file=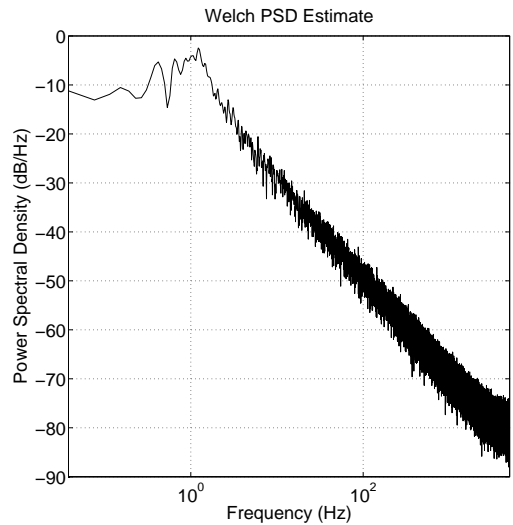,width=7cm, angle=270}
   \caption{Spectrum of the signal shown in Fig.2.}
  \end{center}
\end{figure}

And yet -- there is a simple and clear-cut phase that enters in
this example. The signal shown in Fig.2 is generated by a random 
process defined as 
\be
   x(t) = \cos(\phi(t))
\ee
with the phase performing a biased random walk (cf. \cite{rossberg}),
\be
   d \phi(t)/dt = \omega + \eta(t)
\ee
where $\eta(t)$ is $\delta$-correlated white noise,
\be
   \langle \eta(t) \rangle = 0, \qquad \langle \eta(t)\eta(t') \rangle = D 
\delta(t-t').
\ee
The parameter values used in Figs.2 to 5 are $\omega=1$ Hz and $D=5$, 
and the integration was made with step $\delta t = 0.0001$.
The delay used in Fig.3 was $\tau = 0.015$.

If the noise variance $D$ were much smaller, we would not have 
much problems. The problems arise since we chose a rather
large $D$ such that the phase is not monotonically increasing.
Instead there are large intervals during which the phase decreases,
leading to ``fake" loops in Figs.~3 and 4. Our point is {\it not} that 
presently popular methods for extracting phases cannot distinguish between 
such phase reversals (or even just sudden slow-downs of the instantaneous
phase velocity) and ``true" amplitude variations \cite{footnote}. Rather,
we want to
stress that there is no way {\it in principle} to distinguish between 
them. Thus attempts to improve on phase extraction methods in 
similarly ambiguous situations \cite{rossberg,rosenbl} are likely to lead
to ambiguous results, even if this ambiguity might be hidden.

Ways out of avoiding these ambiguities can be found only by restricting what
we accept as a sensible phase definition. One could argue e.g. that a basic
intuitive feature of a phase is its {\it continuous temporal progression},
i.e. positivity of the instantaneous phase velocity. 
Demanding this would mean that there is no possibility at all to define a
phase for the above model, and the same would be true for a large class 
of signals. 

Does this mean that such a requirement is too restrictive to 
be useful? We believe not. One traditional way out of the dilemma when 
phases should be defined for arbitrary signals is Fourier analysis. One 
decomposes the signal into harmonic components, and can then define phases
for each component (or, when the signal is decomposed into frequency bands, 
for each band). What we propose is to decompose signals more generally
into components with {\it positive} but not necessarily constant
(as in a Fourier decomposition) phase velocities. This added
freedom might allow much more physically relevant decompositions. Indeed, 
we do not have to invent any new example for this, since the best example
demonstrating the power of such an approach is known since nearly four 
hundred years: progress in understanding planetary motion was only possible
when Kepler replaced the decomposition into the harmonic epicycles of Ptolemaeus
and Copernicus by a decomposition into elliptic motions, which are just 
of the type advocated by us \cite{foot2}. 
Details of such a decomposition will of course depend on the problem at 
hand, and we cannot give any general algorithm. But the possibility and 
the eventual usefulness of such an approach should be kept in mind.


\begin{thebibliography}{99}
\bibitem{footnote} A hint at the actual structure of the present time sequence
   is obtained from the fact that it has many degenerate extrema: while 
   there are many local maxima and minima with fluctuating amplitudes,
   the absolute extrema are all at $x(t)=\pm 1$. One might use such
   kind of information in special cases like the present one, but this 
   will hardly lead to a universal and robust algorithm.
\bibitem{foot2} A general ansatz with monotonically increasing phases in the 
   spirit of the above discussion would be $x(t) = f(\phi_1(t),\ldots,\phi_n(t))$ 
   with $\dot{\phi}_i(t)>0$ for $i=1,\ldots,n$.
\end{thebibliography}
\end{document}